\documentclass[preprint,aps,nofootinbib,showpacs,showkeys,12pt]{revtex4}

\usepackage{graphicx}
\usepackage{graphics}
\usepackage{epsfig}

\usepackage{axodraw}
\usepackage{pstricks}
\def\bye{\end{document}}

\begin{document}
\title{{\bf\Large Charged pions and kaons weak leptonic decays at finite density}}
\author{\bf P. Costa}
\email{pcosta@teor.fis.uc.pt}
\author{\bf M. C. Ruivo}
\email{maria@teor.fis.uc.pt}
\affiliation{Departamento de F\'{\i}sica, Universidade de Coimbra,
P-3004-516 Coimbra, Portugal} 
\author{\bf Yu. L. Kalinovsky}
\email{kalinov@nusun.jinr.ru}
\affiliation{Laboratory of Information Technologies,
Joint Institute for Nuclear Research, Dubna, Russia}
\date{\today}
\begin{abstract}
The weak leptonic decays $\pi^\pm\rightarrow\mu^\pm\nu_{\mu}(\bar{\nu}_{\mu}),\,\pi^\pm\rightarrow e^\pm\nu_e(\bar{\nu}_{e}),\,K^\pm\rightarrow\mu^\pm\nu_{\mu}(\bar{\nu}_{\mu})$ and $K^\pm\rightarrow e^\pm \nu_e(\bar{\nu}_{e})$ are investigated within the framework of the $SU(3)$ Nambu-Jona-Lasinio model, in the vacuum and in quark matter in $\beta$ equilibrium. It is found that, at low densities, the decay rates of the charge multiplets, that are decoupled in this kind of flavor asymmetric matter, are affected differently. As the density increases, these decay rates become closer and show a tendency to vanish. This shows that weak interactions are strongly influenced by the medium as a consequence of restoration of chiral symmetry.
\end{abstract}
\maketitle


\section{Introduction}

It is a well known fact that hadrons are supposed to change their properties as a consequence of phase transitions of the QCD vacuum that are expected to occur at high densities and temperatures.  Theoretical and experimental efforts  have been dedicated to heavy-ion physics looking for signatures of the quark gluon plasma, a state of matter with deconfined quarks and restoration of symmetries  \cite{kanaya}. In particular, modifications in the   mass spectrum and decays of pseudoscalar mesons   has  attracted a lot of interest. Since,  in the chiral limit, those mesons are the Goldstone bosons associated with spontaneous chiral symmetry breaking,  their in medium properties  are expected to carry relevant information concerning the restoration of chiral symmetry. 

Several studies on the electromagnetic decays of neutral pseudoscalar mesons pion and $\eta$, at finite density or temperature \cite{tdavid,tklev,pisarski,costag} are known, but less attention has been given to the study of the weak decays of charged mesons, such as $\pi^\pm\,, K^\pm$ \cite{tklev}.  
Although chiral symmetry does not depend on the existence of weak interactions, the symmetry currents $V^\mu$ and $A^\mu$  happen to enter in the strangeness conserving semileptonic weak decay interactions. On the other side, the pion (kaon) decay constants, that can be measured from the weak decays $\pi^+ \rightarrow \mu^+ \nu_u$  ($K^+ \rightarrow \mu^+ \nu_u$),  are quantities strongly influenced by the restoration of chiral symmetry. Since  the decay rates of charged pions and kaons are dominated by $f_\pi\,, m_\pi \,,(f_K\,, m_K )$,   those decays are expected be significantly modified by the medium. 

Much attention is being given nowadays to the investigation of  phase transitions and of hadron properties in  flavor asymmetric matter, in particular neutron matter in weak equilibrium, which is specially relevant for neutron stars. In such asymmetric media,   
the charge multiplets of mesons, are expected to have a splitting  \cite{kaplan,RSP,costaI}. 
The study of weak leptonic decays of charged pions and kaons in  flavor asymmetric media could give interesting information concerning the different modifications produced by the medium on the charge multiplets of mesons. We will address this problem within the $SU(3)$ Nambu- Jona-Lasinio model [NJL].

The [NJL] \cite{NJL}  model is an effective quark model that incorporates important symmetries of QCD and  the mechanisms of symmetry breaking. [NJL] type models have been extensively used
over the past years to describe low energy features of hadrons and also to  investigate restoration of chiral symmetry with temperature or density \cite{njlT,Hatsuda94,RuivoSousa,RSP,buballa}.
The behavior of $SU(3)$ pseudoscalar mesons, in NJL models,  with temperature have been studied in \cite{Hatsuda94,RKS}. The combined effect of temperature and density has investigated in \cite{costaB,costabig}. Different studies have been devoted to the
behavior of pions and kaons at finite density in flavor symmetric
or asymmetric matter \cite{RuivoSousa,RSP,costaI}. Recently, we studied the behavior of two photon decay observables for the neutral mesons $\pi^0\,,\eta$ in quark matter in weak equilibrium and we  have shown that these anomalous decays are strongly influenced by the medium, its amplitudes decreasing with density, which is mainly   a manifestation of restoration of chiral symmetry \cite{costag}.  The aim of the present work is to study the weak decays: $\pi^\pm\rightarrow\mu^\pm\nu_{\mu}(\bar{\nu}_{\mu}),\,\pi^\pm\rightarrow e^\pm\nu_e(\bar{\nu}_{e}),\,K^\pm\rightarrow\mu^\pm\nu_{\mu}(\bar{\nu}_{\mu})$ and $K^\pm\rightarrow e^\pm \nu_e(\bar{\nu}_{e})$ in order to discuss the effects on these decays of the restoration of chiral symmetry. We start by giving a brief description of general formalism in section 2, we give a theoretical background of the weak decays in section 3 and in section 4 we present and discuss our  results.


\section{Formalism}

We consider the three--flavor  NJL type model containing  
scalar--pseudoscalar interactions and a determinantal term, the 't Hooft
interaction which breaks the $U_A(1)$ symmetry. The model has
the following Lagrangian: 

\begin{equation}
\begin{array}{rcl}
{\cal L\,} & = & \bar q\,(\,i\, {\gamma}^{\mu}\,\partial_\mu\,-\,\hat m)\, q+%
\frac{1}{2}\,g_S\,\,\sum^8_{a=0}\, [\,{(\,\bar q\,\lambda^a\, q\,)}%
^2\,\,+\,\,{(\,\bar q \,i\,\gamma_5\,\lambda^a\, q\,)}^2\,] \\[4pt] 
& + & g_D\,\, \{\mbox{det}\,[\bar q\,(\,1\,+\,\gamma_5\,)\,q\,] + \mbox{det}%
\,[\bar q\,(\,1\,-\,\gamma_5\,)\,q \,]\, \}. \label{1} \\ 
&  & 
\end{array}
\label{lag}
\end{equation}
Here $q = (u,d,s)$ is the quark field with three flavors, $N_f=3$, and
three colors, $N_c=3$. $\hat{m}=\mbox{diag}(m_u,m_d,m_s)$ is the current 
quark masses matrix and $\lambda^a$ are the Gell--Mann matrices, 
a = $0,1,\ldots , 8$, ${ \lambda^0=\sqrt{\frac{2}{3}} \, {\bf I}}$.

By using standard bosonization techniques one obtains the meson effective action  (for details see \cite{costabig,costaB} and references therein):
\begin{eqnarray}  \label{action}
W_{eff}=&-i& \mbox{Tr} \ {\rm ln}(\,i\,\partial_\mu \gamma_{\mu}-\hat
m+\sigma_a\,\lambda^a +i\,\gamma_5\, {\phi}_a\,\,{\lambda}^a\,)  \nonumber \\
&-&\frac{1}{2}(\,\sigma_a\,{S_{ab}}^{-1}\,\sigma_b\,+{\phi}_a\, {\ P_{ab}}%
^{-1}\,\phi_b\,).
\end{eqnarray}

Where the fields $\sigma^a$ and $\phi^a$  are the  scalar and pseudoscalar 
meson nonets 
and $S_{ab}\,=\,g_S\,\delta_{ab}\,+\,g_D D_{abc} \,< \bar q\, {\lambda}^c\,q\,>,$
$P_{ab}\,=\,g_S\,\delta_{ab}\,-\,g_D D_{abc} \,< \bar q\, {\lambda}^c\,q\,>$ are projectors. The constants $D_{abc}$ are  the $SU(3)$  structure constants,  for $a,b,c  \in \{1,2,..8\}\,$,  and $D_{000}=\sqrt{\frac{2}{3}}$,
$D_{0ab}=-\sqrt{\frac{1}{6}}\delta_{ab}$.
The first variation of the action (\ref{action}) leads to the usual set of 
gap equations for constituent quark masses $M_i$ and   the pseudoscalar meson masses,  $M_H$ ($H =\pi^{\pm},\,K^{\pm}$), where the polarization operator $\Pi_{ab}(P)$ is  
given by:
\begin{eqnarray}\label{polop}
\Pi_{ab} (P) = i N_c \int \frac{d^4p}{(2\pi)^4}\mbox{tr}_{D}\left[
S_i (p) (\lambda^a)_{ij} (i \gamma_5 )
S_j (p+P)(\lambda^b)_{ji} (i \gamma_5 )
\right],
\end{eqnarray}
with $S_i (p)$ being the quark propagator.
The quark--meson coupling constant $g_{H\bar{q}q}$ and the 
meson decay constant $f_H$  are defined as usual by the expressions:  
\begin{equation}
g_{H\bar{q}q}^{-2}=%
-\frac{1}{2 M_H}\frac{\partial}{\partial P_0}[\Pi_{ab}(P_0)]_{|P_0=M_H},
\label{coupl}
\end{equation}

\begin{eqnarray}
f_H = N_c g_{H\bar{q}q} \frac{P_\mu}{P^2} 
\int \frac{d^4 p}{(2\pi)^4}
\mbox{tr} \left[ (i \gamma_5) S_i(p) (\gamma_\mu \gamma_5) S_j(p+P) \right].
\end{eqnarray}

The model parameters  are  fitted to the experimental values of masses for
pseudoscalar mesons ($M_{\pi} = 139$ MeV, $M_{K} = 495.7$) and 
$f_\pi = 93.0$ MeV\footnote{For each meson, our definition of the meson decay constant $f_H$
differs by a factor $1/\sqrt2$ from the one accepted by the Particle Data Group an given in \cite{pdb}.}. We use the following  parameterization \cite{RSP,RuivoSousa}: 
$\Lambda=631.4$ MeV, 
$g_S \Lambda^2=3.658$, 
$g_D\Lambda^5=-9.40$, 
$m_u=m_d=5.6$ MeV and 
$m_s=135.6$ MeV. For the quark condensates we have: 
$<\bar{u}u>=<\bar{d}d>=-(246.7 \mbox{MeV})^3$, 
$<\bar{s}s>=-(266.9 \mbox{MeV})^3$, and 
$M_u = M_d = 335.4$ MeV, $M_s = 527.4$ MeV, for the constituent       
quark masses. 


\section{$\pi^{\pm}$ and $K^{\pm}$ leptonic decays }

The standard form of the invariant amplitude for the weak decay  of kaons and pions, schematically represented in Fig. 1, is:  
\begin{equation}
\mathcal{M} = \frac{{\tilde{G}}}{\sqrt{2}}  f_{H}\, q^{\rho} (\overline{u}(p)\gamma _{\rho }(1-\gamma
_{5}) v(k))\,,
\end{equation}
\noindent where   $f_{H}$ stands for the pion (kaon) decay constant,   $u(p)\,,v(k)$ are  the spinors for the muon and the neutrino, respectively; the term in brackets is the leptonic current and $f_H q^{\rho}$ represents the effective current of bound quarks inside the meson. For the kaon ${\tilde{G}}\rightarrow G \sin \theta_{C}$ and for the pion $\tilde{G}\rightarrow G \cos\theta_{C}$, where
 $G=1.16639(1)\times10^{-5}$GeV$^{-2}$ is the Fermi constant \cite{pdb} and $\theta_{C}=13^{0}$ is the Cabibbo angle \cite{Halzen,bernstein}.

\begin{figure}[t]
\[
\begin{picture}(125,86)(80,80)
\hspace*{2cm}
\SetScale{2.0}
{
\ArrowLine(0,30)(35,30)
\ArrowLine(35,30)(60,60)
\ArrowLine(35,30)(60,0)
\Vertex(35,30){2}
\Text(30,45)[lb]{$q$}
\Text(80,90)[lb]{$p$}
\Text(80,20)[lb]{$k$}
\Text(0,70)[lb]{\large $\pi^+$, $K^+$}
\Text(125,120)[lb]{\large $\mu^+$}
\Text(125,-10)[lb]{\large $\nu_{\mu}$}
}
\end{picture}
\]
\vspace{2.5cm}\caption{Feynman diagram for the decay
$\pi^+,K^+(q)\rightarrow\mu^+(p)+\nu_{\mu}(k)$ with four-momentum $q=p+k$.}
\label{fig:direct1}
\end{figure}
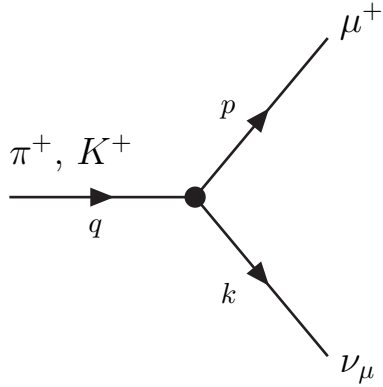
The decay rate is given by:
\begin{equation}
d\Gamma = \frac{1}{2 m_{H}} \overline{{|\cal M|}^2} \frac{d^3p}{ (2 \pi)^3 2 E} \frac{d^3k}{ (2 \pi)^3 2 \omega}\,\delta (q-p-k)\,,
\end{equation}
\noindent where $E, \omega$ are the energies of the muon and of the neutrino, respectively. After calculating the trace and performing phase space integration, one gets, for the decays of pions and kaons:

\begin{equation}
\Gamma _{(\pi ^{\pm }\rightarrow \mu ^{\pm }\nu _{\mu }(\bar{\nu}_{\mu }))}=%
\frac{(G\cos\theta_c)^{2}}{4\pi }f_{\pi ^{\pm }}^{2}m_{\pi ^{\pm }}m_{\mu }^{2}\left( 1-%
\frac{m_{\mu }^{2}}{m_{\pi ^{\pm }}^{2}}\right) ^{2}
\end{equation}

\begin{equation}
\Gamma _{(K^{\pm }\rightarrow \mu ^{\pm }\nu _{\mu }(\bar{\nu}_{\mu }))}=%
\frac{(G \sin \theta _{c})^{2}}{4\pi }f_{K^{\pm }}^{2}m_{K^{\pm }}m_{\mu
}^{2}\left( 1-\frac{m_{\mu }^{2}}{m_{K^{\pm }}^{2}}\right) ^{2}
\end{equation}
Similar expressions are obtained for the decays $(\pi ^{\pm }\rightarrow e^{\pm }\nu _{e}(\bar{\nu}_{e}))$ and $(K^{\pm }\rightarrow e^{\pm }\nu _{e}(\bar{\nu}_{e}))$, with $m_{\mu} \rightarrow m_e$. 

The ratios of the rates of the decays mentioned above are easily calculated through the expressions:

\begin{equation}
R_{W}=\frac{\Gamma _{(\pi ^{\pm }\rightarrow e^{\pm }\nu _{e}(\bar{\nu}_{e}))}}{\Gamma _{(\pi ^{\pm }\rightarrow \mu ^{\pm }\nu _{\mu }(\bar{\nu}_{\mu
}))}}%
=\left( \frac{m_{e}}{m_{\mu }}\right) ^{2}\left( \frac{m_{\pi ^{\pm
}}^{2}-m_{e}^{2}}{m_{\pi ^{\pm }}^{2}-m_{\mu }^{2}}\right) ^{2};
\end{equation}
  
\begin{equation}
R_{W}=\frac{\Gamma _{(K^{\pm }\rightarrow e^{\pm }\nu _{e}(\bar{\nu}_{e}))}}{\Gamma _{(K^{\pm }\rightarrow \mu ^{\pm }\nu _{\mu }(\bar{\nu}_{\mu}))}}%
=\left( \frac{m_{e}}{m_{\mu }}\right) ^{2}\left( \frac{m_{K^{\pm
}}^{2}-m_{e}^{2}}{m_{K^{\pm }}^{2}-m_{\mu }^{2}}\right) ^{2}.
\end{equation}


It is an interesting feature that the branching ratio  for the leptonic decays 
of pions (kaons) into muons be much larger than the energetically favored 
decays in the electron channel. In fact, the branching ratios for 
$\pi ^{+}\rightarrow \mu ^{+}\nu _{\mu }\,\mbox{is}\, 99.987\%;$, while  for $\pi ^{+}\rightarrow e ^{+}\nu _{e }\, \mbox{is only}\, (1.230 \pm 0.004) \times 10^{-4}\%$.
The branching ratios for kaons show the same tendency: $K^{+}\rightarrow \mu ^{+}\nu _{\mu }\,\, \mbox{is}\, \, 63.43\pm0.17 $ and $K^{+}\rightarrow e^{+}\nu _{e}\, \mbox{is}\,  (1.55\pm0.07)\times 10^{-5}\%$. This due to the helicity suppression phenomena. In fact, the weak interaction current contains the left-handed projection operator which, in the massless limit, produces only left-handed particles and right-handed antiparticles. However, due  to angular momentum conservation, the decay of a spin zero particle to massless $e^{\pm }\nu _{e}(\bar{\nu}_{e})\,, \mu ^{\pm}\nu_{\mu}(\bar{\nu}_{\mu })$ is forbidden because the lepton would require to have combined angular momentum $J_z=1$. Thus the amplitude for the leptonic decays would vanish for $m_{\mu}=m_{e}=0$. But when $m_{\mu}\not=0\, (m_{e}\not=0)$, the muon (electron) is  in a state of mixed helicities, which allows the decay.  Since the coefficient of the positive helicity state in the mixed state is proportional to the muon (electron) mass, it is clear that a decay into a state of higher mass is prefered; similar considerations apply to the kaon decays \cite{bernstein}. It can also be noticed  (see Table I) that the ratios of the decay rates for kaons are lower by an order of magnitude relative to pions. This is because, due to the larger  mass of the kaon, as compared to that of pion, a stronger helicity suppression of the electron mode relative to the muon mode is required. This  argument leads us to expect that the amplitudes of leptonic decays of pions and kaons in the medium decrease considerably, since the masses of these mesons increase with density.


\section{Discussion and conclusions} 

We present in Table I our results for 
the decay rates and ratios involved in the decays $\pi^\pm\rightarrow\mu^\pm\nu_{\mu}(\bar{\nu}_{\mu}),\,\pi^\pm\rightarrow e^\pm\nu_e(\bar{\nu}_{e}),\,K^\pm\rightarrow\mu^\pm\nu_{\mu}(\bar{\nu}_{\mu})$ and $K^\pm\rightarrow e^\pm \nu_e(\bar{\nu}_{e})$ in the vacuum, in comparison with experimental results. We can see that there is a good overall agreement; the reason for the value of the kaons width being not so good is due to the fact that the value for $f_K$ in the NJL model is lower than the experimental value (exp: 113 MeV, NJL 97.7 MeV) and $f_K$ enters explicitly in the expression for this width (see Eq. 10).

\begin{table}[t]
\caption{Comparison of the results for the amplitude $\Gamma$ and $R_W$
obtained in the NJL model with experimental values.}
\begin{tabular}
[c]{cc||c|c|}\cline{3-4}%
&  & NJL & Exp.\\\hline\hline
\multicolumn{1}{|c}{$\Gamma$ [MeV]} & \multicolumn{1}{|c||}{$\pi
^{+}\rightarrow\mu^{+}\nu_{\mu}$} & $2.50\times10^{-14}$ & $2.52\times
10^{-14}$\\[0.2cm]\cline{2-4}\cline{4-4}%
\multicolumn{1}{|c}{} & \multicolumn{1}{|c||}{$K^{+}\rightarrow\mu^{+}\nu
_{\mu}$} & $2.64\times10^{-14}$ & $3.37\times10^{-14}$\\[0.2cm]\hline%
\multicolumn{1}{|c}{$R_{W}$} & \multicolumn{1}{|c||}{$\frac{\Gamma{(\pi
^{+}\rightarrow e^{+}\nu_{e})}}{\Gamma{(\pi^{+}\rightarrow \mu^{+}\nu_{\mu})}}%
$} & $1.23\times10^{-4}$ & $1.23\times10^{-4}$\\[0.2cm]\cline{2-4}%
\multicolumn{1}{|c}{} & \multicolumn{1}{|c||}{$\frac{\Gamma{(K
^{+}\rightarrow e^{+}\nu_{e})}}{\Gamma{(K^{+}\rightarrow \mu^{+}\nu_{\mu})}}%
$} & $2.57\times10^{-5}$ & $2.45\times10^{-5}$\\[0.2cm]\cline{1-4}%
%
\end{tabular}
\end{table}

Now let us discuss our results at finite density. We  consider here the case of asymmetric quark matter imposing 
the condition of $\beta$ equilibrium  and charge neutrality through the following 
constraints, respectively on the chemical potentials and densities of quarks and electrons: $\mu_{d}=\mu_{s}=\mu_{u}+\mu_{e}\,\, \mbox{ and }\,\,\,\frac{2}{3}\rho_{u}
-\frac{1}{3}(\rho_{d}+\rho_{s})-\rho_{e}=0$, with 
$\rho_{i}=\frac{1}{\pi^{2}}(\mu_{i}^{2}-M_{i}^{2})^{3/2}\theta(%
\mu_{i}^{2}-M_{i}^{2})\,\,\mbox{ and }\,\,\,\rho_{e}=\frac{\mu_{e}^{3}}{3\pi^{2}}$ 
\cite{costaI,costaB,costabig}.

As it has been discussed by several authors, this  version of the NJL model exhibits a first order 
phase transition \cite{RSP,costaI,buballa}. An interesting feature of quark matter in weak equilibrium  is that at densities above $\rho_s\sim 3.9\rho_0$ the mass of the strange quark becomes  lower than the chemical potential  what implies the occurrence   of strange valence quarks in this regime.  This fact   leads  to meaningful effects on the behavior of meson observables \cite{costaI,costaB,costabig}.

In order to understand our results for $\Gamma$ and $R_W$ as functions of the density, we plot first (see Fig. 2) the masses of the charged kaons and pions as well as its decays constants as functions of the density, and we   can see the  
splitting between charge multiplets (Fig. 2 a)): the increase of the masses of $K^+ $ and $\pi^- $
with respect to those of $K^-$ and $\pi^+$ is due to Fermi blocking.
 At the critical density the antikaons enter in the continuum, but they become again bound states above $\rho_0 \sim 4 \rho_0$. 
\begin{figure}[t]
    \begin{center}
			a)\epsfig{file=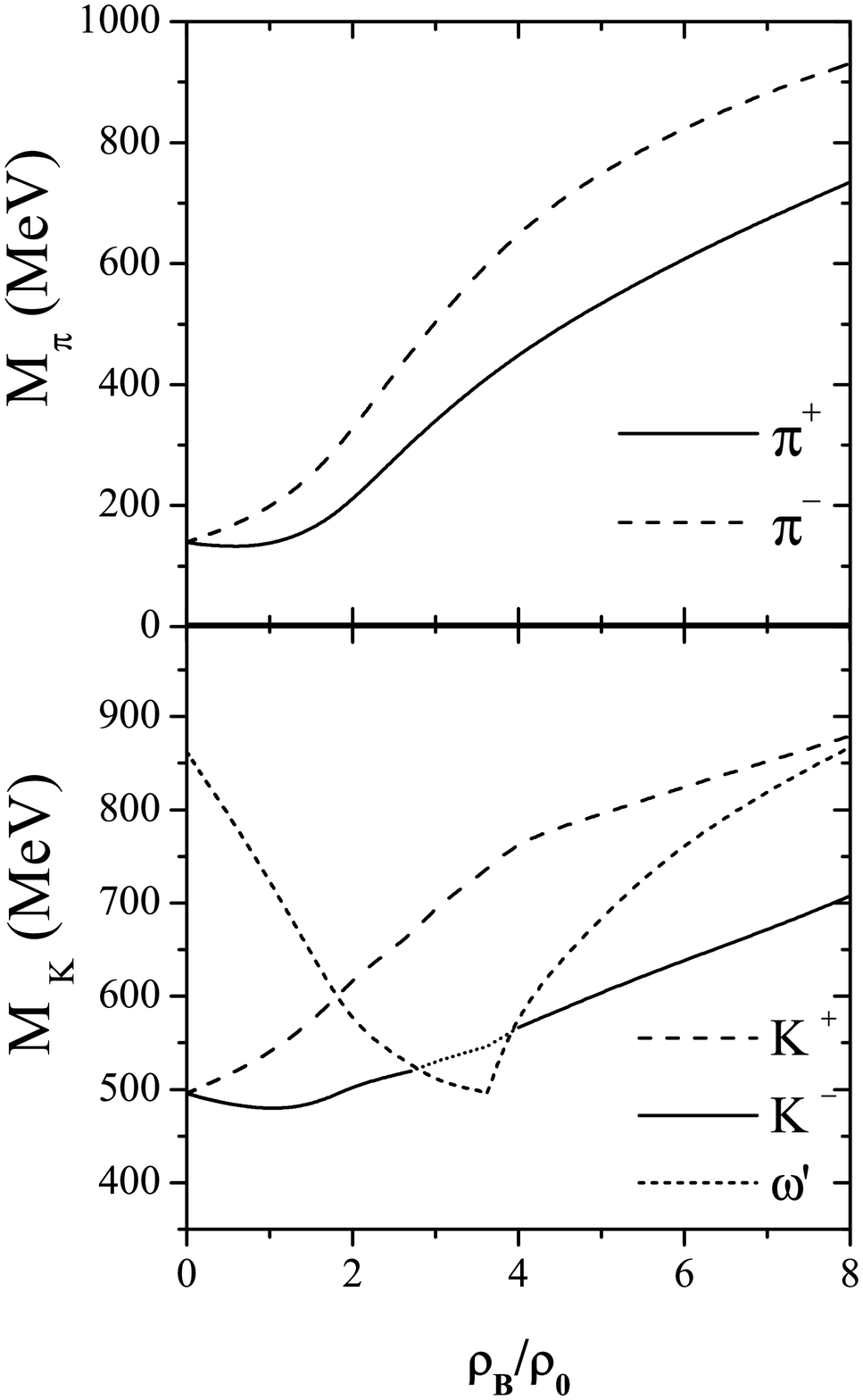, width=6.5cm,height=9cm}
			b)\epsfig{file=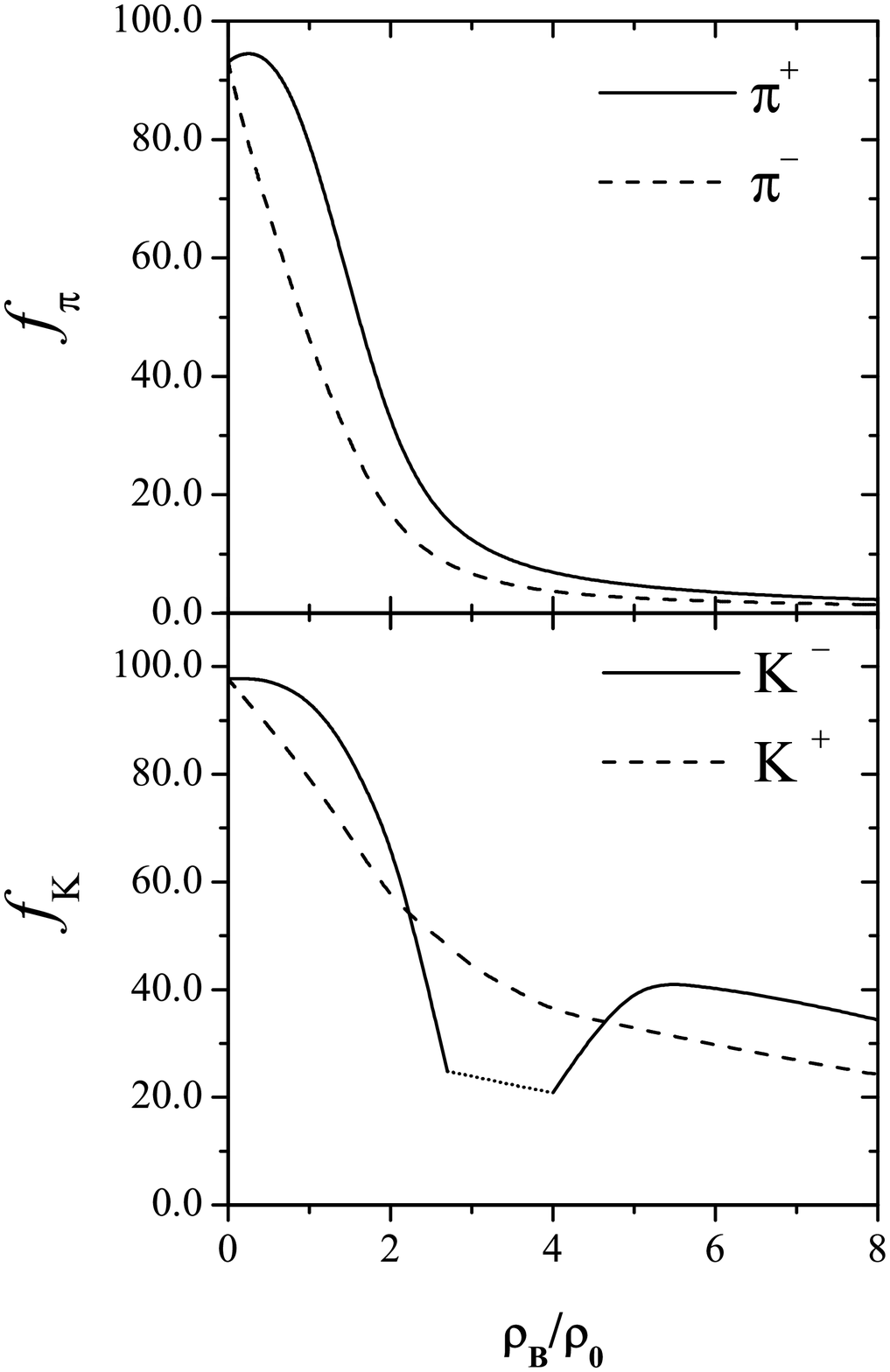, width=6.5cm,height=9cm}
    \end{center}
		\caption{Masses a) and decay constants b) of charged pions and kaons as functions of the  			density.}
\label{fig2}
\end{figure}
As for the pion decay constants (Fig. 2 b)), although there is a splitting for moderate densities, as the density increase they go assimptotically to zero, which is an indication of the restoration of chiral symmetry in the non-strange sector. The decrease is more moderate for the kaon decay constant, which reflects the fact that chiral symmetry is weakly restored in the strange sector. The   region of densities ($\sim 2.5 - 4 \rho_0$) where the $K^-$ decay constant is flat corresponds to the region where this meson lies in the continuum.

Concerning the decay rate 
$\Gamma (\pi \rightarrow\,\mu\nu_{\mu})$,  a similar behavior to the observed with temperature \cite{tklev} is expected  with density, but, since in the medium considered here there is flavor asymmetry,  there is a meaningful difference between the behavior of $\pi^+ \mbox{and } \pi^-$ at low densities (see Fig. 3 a), upper panel), which is due the interplay between the behavior of the pion masses and decay constants. As the density increases, the two decay constants become closer, and goes to zero and, since $f_\pi $ has a dominant effect,  the decay rates become almost  degenerated and vanishing. A similar behavior is found for the charged kaons, although less pronounced, as expected (Fig. 3 a), lower panel). We notice that, as the density increases, on one side, due to the increase of the meson masses, the phase space available  ($\propto  ({m_{H^{\pm }}}^{2}-m_{\mu }^{2})^{2}\,, ({m_{H^{\pm}}}^{2}-m_{e}^{2})^{2})$ increases; on the other side $f_{H}$ decreases and this is the dominant effect in the decay rates. Notice that the current of quarks bound inside the mesons is represented by $f_{H} q^{\rho}$ (see Eq. (7)), therefore the decrease of $f_{H}$ is an indication that the quark current becomes weaker, meaning that the mesons became less quark bound states.
\begin{figure}[t]
    \begin{center}
			a)\epsfig{file=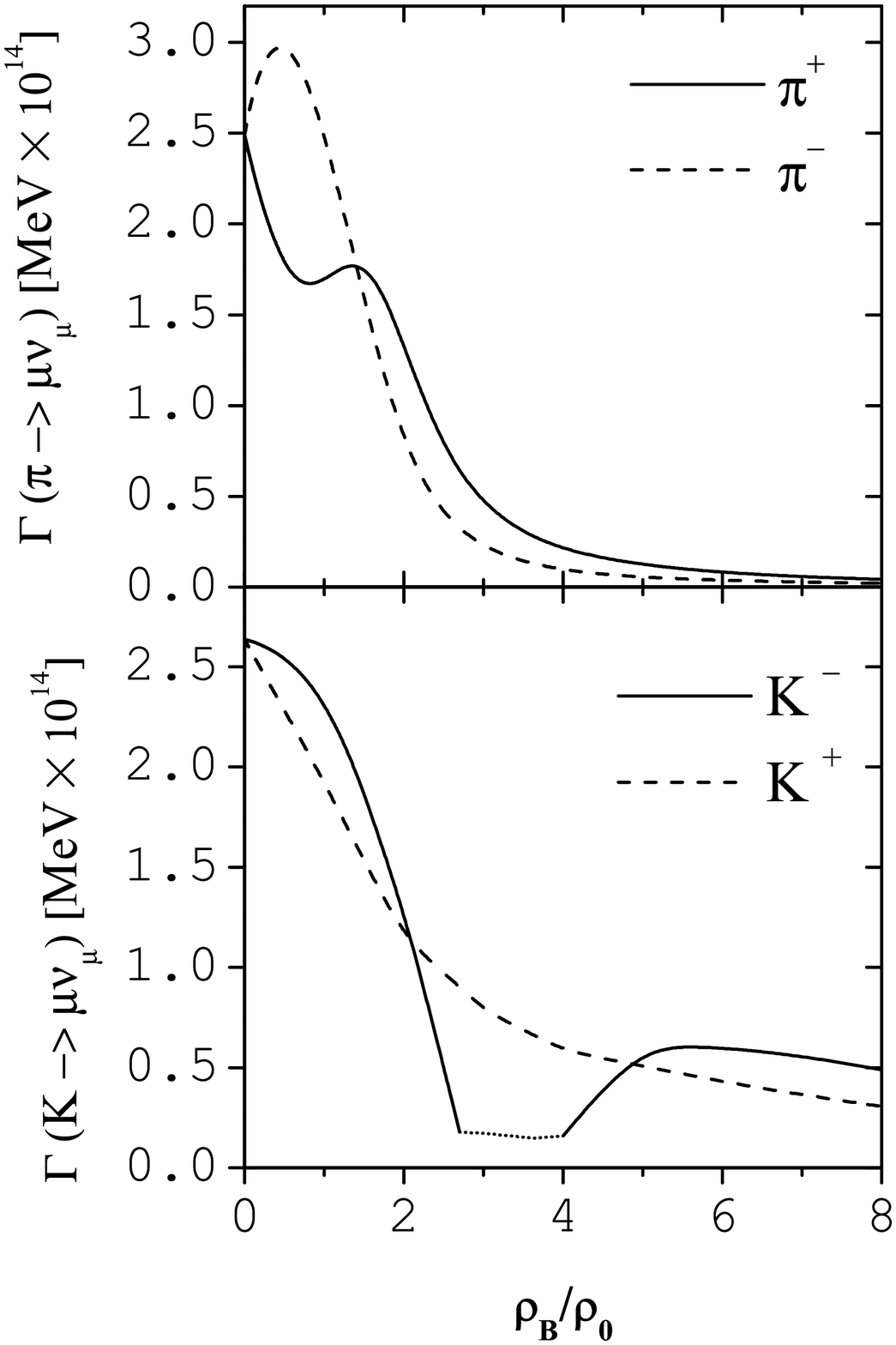, width=6.5cm,height=9cm}
			b)\epsfig{file=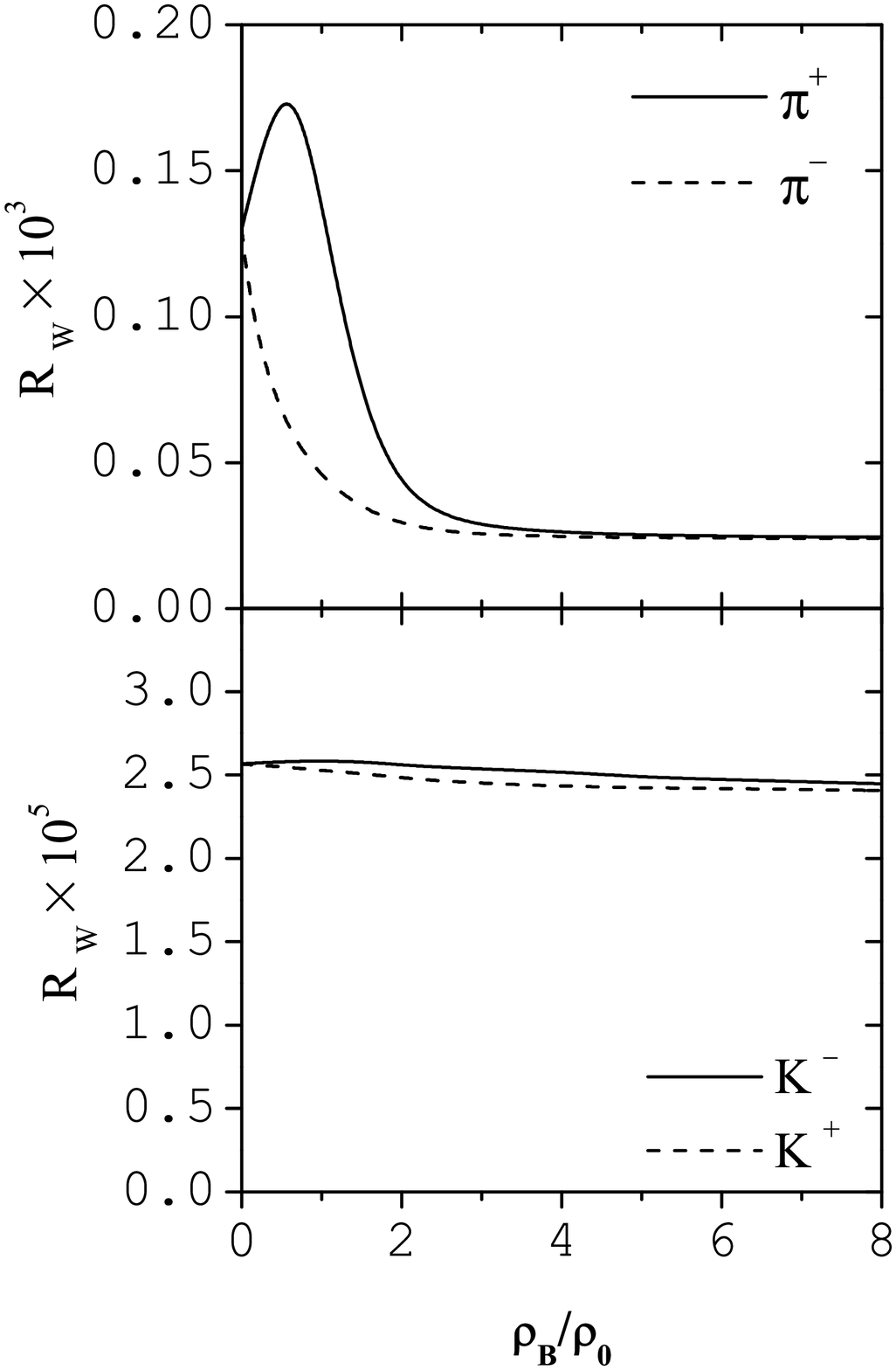, width=6.5cm,height=9cm}
    \end{center}
    \caption{$\Gamma$ a) and $R_W$ b) for charged pions and kaons as functions of density.}
    \label{fig3}
\end{figure}

Concerning  the ratios, $R_W$,  in the vacuum they are very small, due to helicity suppression as explained before. In the medium, since there is a splitting between the mesons masses,   the phase space available is different for different charge multiplets and, therefore we observe a splitting in $R_W$ (Fig. 3b)). But as the density increases the phase space available for the decay into  muons becomes closer to the one for the decay into electrons: $ ({m_{H^{\pm }}}^{2}-m_{\mu }^{2})^{2}\approx({m_{H^{\pm}}}^{2}-m_{e}^{2})^{2}$ due to the increase of meson masses.

In conclusion, one can say that  weak interactions, as already was shown for anomalous interaction \cite{costag}, are strongly influenced by the medium and reflect the restoration of chiral symmetry.


\vspace{2cm}
\begin{center}
{\large Acknowledgment:}
\end{center}
Work supported by grant SFRH/BD/3296/2000 (P. Costa), by grant RFBR 03-01-00657, 
Centro de F\'{\i}sica Te\'orica and GTAE (Yu. Kalinovsky).



\end{document}